# Not So Easy Problems for Tree Decomposable Graphs[*]

Stefan Szeider

Institute of Information Systems
Vienna University of Technology
A-1040 Vienna, Austria
stefan@szeider.net

**Abstract**

We consider combinatorial problems for graphs that (a) can be solved in polynomial time for graphs of bounded treewidth and (b) where the order of the polynomial time bound is expected to depend on the treewidth of the considered graph. First we review some recent results for problems regarding list and equitable colorings, general factors, and generalized satisfiability. Second we establish a new hardness result for the problem of minimizing the maximum weighted outdegree for orientations of edge-weighted graphs of bounded treewidth.

*Keywords:* Treewidth, W[1]-Hardness, Graph Coloring, General Factors, Generalized Satisfiability, Minimum Maximum Outdegree Orientations

## 1 Introduction

*Treewidth* is a graph invariant that indicates, in a certain sense, the global connectivity of a graph. Graphs of treewidth at most $k$ are also known as partial $k$-trees and width-$k$ tree decomposable graphs. Treewidth plays a central role in Robertson and Seymour's Graph Minors Project and has important algorithmic applications. Many hard graph problems are easy for graphs of small treewidth; for example, 3-COLORABILITY and HAMILTONICITY can be solved in linear time for graphs of treewidth bounded by a constant $k$ (albeit with a running time containing a constant factor that is exponential in $k$). In fact, all problems that can be expressed in the formalism of Monadic Second Order Logic (that includes the two mentioned problems, but also linear optimization problems like DOMINATING SET) can be solved in linear time for graphs of bounded treewidth [8, 2]. However, there are problems that are NP-hard for graphs of a certain fixed treewidth bound; for example BANDWIDTH is NP-hard for graphs of treewidth 1 [13] and $L(2,1)$-COLORING is NP-hard for graphs of treewidth 2 [11] (to name an old and a new result).

In this paper we focus on problems that are, in a certain sense, neither very easy nor very hard for graphs of bounded treewidth and thus lie between the two extremes. More specifically, we focus on problems that can be solved in polynomial time for graphs of bounded treewidth, but where the order of the polynomial that bounds the running time necessarily depends on the treewidth bound. The theoretical framework of *parameterized complexity* provides the concepts and methods for providing evidence that a certain problem is of this type. The key method is to show that the problem at hand is W[1]-*hard under fpt-reductions* where W[1] is a complexity class that is considered as the parameterized analog of NP. As NP-hardness provides strong evidence that there is no polynomial-time algorithm for a problem, W[1]-hardness provides strong evidence that a problem cannot be solved in polynomial time for instances of bounded treewidth such that the order of the polynomial is independent of the treewidth bound.

We provide definitions and background information on treewidth and parameterized complexity in Section 2. In Section 3 we review some recent W[1]-hardness result for problems on graphs of bounded treewidth, including problems regarding list and equitable colorings, general factors, and generalized satisfiability. In Section 4 we establish a new W[1]-hardness result for the MINIMUM MAXIMUM OUTDEGREE problem for edge-weighted graphs.

## 2 Preliminaries

### 2.1 Graphs and Tree decompositions

All considered graphs are finite, simple and undirected, unless stated otherwise. We denote the vertex set and the edge set of a graph $G$ by $V(G)$ and $E(G)$, respectively, and an edge between vertices $u$ and $v$ by $uv$ (or equivalently $vu$). Furthermore, we denote the subgraph of a graph $G$ induced by a set $X \subseteq V(G)$ by $G[X]$; that is, $V(G[X]) = X$ and $E(G[X]) = \{uv \in E(G) : u, v \in X\}$. We also write $G - X = G[V(G) \setminus X]$.

A *tree decomposition* of a graph $G$ is a pair $(T, \chi)$ where $T$ is a tree and $\chi$ is a mapping that assigns to each vertex $t \in V(T)$ a set $\chi(t) \subseteq V(G)$ such that the following conditions hold:

1. $V(G) = \bigcup_{t \in V(T)} \chi(t)$ and $E(G) \subseteq \bigcup_{t \in V(T)} \{uv : u, v \in \chi(t)\}$.

2. The sets $\chi(t_1) \setminus \chi(t)$ and $\chi(t_2) \setminus \chi(t) = \emptyset$ are disjoint for any three vertices $t, t_1, t_2 \in V(T)$ such that $t$ lies on a path from $t_1$ to $t_2$ in $T$.

The *width* of $(T, \chi)$ is $\max_{t \in V(T)} |\chi(t)| - 1$. The *treewidth* $tw(G)$ of $G$ is the smallest integer $k$ such that $G$ has a tree decomposition of width $k$. For more information on treewidth we refer to other sources [5, 16].

We shall frequently use the following observation.

**Observation 1.** *Let $G$ be a graph and $X \subseteq V(G)$. Then $tw(G) \leq tw(G - X) + |X|$.*

*Proof.* If $(T, \chi)$ is a tree decomposition of $G - X$, then $(T, \chi')$, with $\chi'(t) = \chi(t) \cup X$ for $t \in V(T)$, is a tree decomposition of $G$. □

It is NP-hard to determine the treewidth of a graph [1]. However, for fixed $k \geq 1$, one can decide in linear time whether the treewidth of a graph is at most $k$, and if so, compute a tree decomposition of width $k$ (Bodlaender's Theorem [4]).

### 2.2 Parameterized Complexity

Let us first review some basic concepts of Parameterized Complexity; for more information we refer to the books of Downey and Fellows [9], Flum and Grohe [12], and Niedermeier [22]. An instance of a parameterized problem is a pair $(x, k)$, where $x$ is the *main part* and $k$ (usually a non-negative integer) is the *parameter*. A parameterized problem is *fixed-parameter tractable* if it can be solved in time $O(f(k)|x|^c)$ where $f$ is a computable function and $c$ is a constant independent of $k$. FPT denotes the class of all fixed-parameter tractable decision problems. A parameterized problem $P$ *fpt-reduces* to a parameterized problem $Q$ if we can transform an instance $(x, k)$ of $P$ into an instance $(x', g(k))$ of $Q$ in time $O(f(k)|x|^c)$ ($f, g$ are arbitrary computable functions, $c$ is a constant) such that $(x, k)$ is a yes-instance of $P$ if and only if $(x', g(k))$ is a yes-instance of $Q$. This definition ensures that if there exists an fpt-reduction from $P$ to $Q$ and $Q$ is fixed-parameter tractable, then so is $P$. A parameterized complexity class $\mathcal{C}$ is the class of parameterized decision problems fpt-reducible to a certain parameterized decision problem $Q_\mathcal{C}$. A parameterized problem $P$ is $\mathcal{C}$-hard if $Q_\mathcal{C}$ (and so each problem in $\mathcal{C}$) can be fpt-reduced to $P$. A $\mathcal{C}$-hard problem that belongs to $\mathcal{C}$ is $\mathcal{C}$-complete. Of particular interest is the class W[1] that is considered as the parameterized analog to NP. It is believed that FPT $\neq$ W[1], and there is strong theoretical evidence that supports this belief; for example, FPT = W[1] implies that the Exponential Time Hypothesis fails (cf. [12]). There are parameterized problems that are believed to be "harder" than problems in W[1]; indeed, there is an infinite hierarchy of parameterized complexity classes FPT = W[0] $\subseteq$ W[1] $\subseteq$ W[2] $\subseteq$ W[3] $\subseteq \cdots$ where all inclusions are believed to be strict.

The following problem is well known to be W[1]-complete [9].

> CLIQUE
>
> *Instance:* A graph $G$ and a non-negative integer $k$.
>
> *Parameter:* The integer $k$.
>
> *Question:* Does $G$ contain a clique on $k$ vertices?

As observed by Pietrzak [23] the problem remains W[1]-complete if the input graph is $k$-partite, which gives the following problem.

> PARTITIONED CLIQUE
>
> *Instance:* A $k$-partite graph $G$ with partition $V_1, \ldots, V_k$ such that $|V_1| = \cdots = |V_k|$.
>
> *Parameter:* The integer $k$.
>
> *Question:* Does $G$ contain a clique on $k$ vertices?

PARTITIONED CLIQUE (also called MULTICOLORED CLIQUE) is particularly useful for reductions in the context of bounded treewidth. Several W[1]-hardness results that we consider in the sequel are obtained by fpt-reductions from this problem.

## 3 Some known W[1]-Hardness Results

### 3.1 Coloring Problems

List coloring is an extensively studied variant of graph coloring [15, 30, 31].

> LIST COLORING
>
> *Instance:* A graph $G$ and for each vertex $v \in V(G)$ a list $l(v)$ of allowed colors for $v$.
>
> *Question:* Is there a proper coloring for $G$ where each vertex is colored with a color from its list?

**Theorem 1** ([10]). LIST COLORING *is* W[1]*-hard when parameterized by the treewidth of the instance graph.*

We sketch the proof as it is very simple and provides a good example for reductions from PARTITIONED CLIQUE.

Consider a $k$-partite graph $G$ with partition $V_1, \ldots, V_k$. We construct a graph $H$ as follows. Let $b : V(G) \to \{1, \ldots, |V(G)|\}$ be an arbitrary but fixed bijection. First we take new vertices $v_1, \ldots, v_k$, and set $l(v_i) = \{b(v) : v \in V_i\}$ ($1 \leq i \leq k$). Second, for all $1 \leq i < j \leq k$ and each pair of *non*adjacent vertices $u \in V_i$, $v \in V_j$ we add a vertex $v_{uv}$ and make it adjacent with $v_i$ and $v_j$; we put $l(v_{uv}) = \{b(u), b(v)\}$. It is easy to verify that $H$ has a proper list coloring if and only if $G$ has a clique on $k$ vertices. Note that $H - \{v_1, \ldots, v_k\}$ is edge-less and so of treewidth 1. Thus $\mathrm{tw}(G) \leq k + 1$ follows by Observation 1. So there is indeed an fpt-reduction from PARTITIONED CLIQUE to LIST COLORING parameterized by the treewidth of the instance graph.

Let us briefly mention a fixed-parameter tractability result that contrasts Theorem 1. A graph $G$ is called *r-list-colorable* or *r-choosable* if for every list assignment $l$ such that $|l(v)| \geq r$ for each vertex $v \in V(G)$, there exists a proper coloring for $G$ where each vertex is colored with a color from its list. The *list-chromatic number* or *choice number* of $G$ is the smallest integer $r$ such that $G$ is *r-list-colorable*. Now, as shown by Fellows et al. [10], determining the list chromatic number of a given graph is fixed-parameter tractable when parameterized by the treewidth of the graph.

Consider the following problem.

> PRECOLORING EXTENSION
>
> *Instance:* A graph $G$ and a proper coloring $c'$ of some induced subgraph $G'$ of $G$ using colors from $\{1, \ldots, r\}$.
>
> *Question:* Is it possible to extend $c'$ to a proper coloring $c$ of $G$ using only colors from $\{1, \ldots, r\}$?

One can fpt-reduce LIST COLORING to PRECOLORING EXTENSION by encoding the lists by means of precolored vertices of degree one, without increasing the treewidth.

**Corollary 1** ([10])**.** PRECOLORING EXTENSION *is* W[1]-*hard when parameterized by the treewidth of the instance graph.*

The next problem was introduced by Meyer [21] motivated by a garbage truck scheduling problem; for history and recent results see [6, 18].

> EQUITABLE COLORING
>
> *Instance:* A graph $G$ and a positive integer $r$.
>
> *Question:* Is there a proper coloring of $G$ using colors from $\{1, \ldots, r\}$ such that the sizes of any two color classes differ at most by one?

**Theorem 2** ([10])**.** EQUITABLE COLORING *is* W[1]-*hard when parameterized by the treewidth of the instance graph. The problem remains* W[1]-*hard when we parameterize simultaneously by the treewidth and the number $r$ of colors.*

Theorem 2 can be shown by a reduction from PARTITIONED CLIQUE; the reduction is significantly more complicated than the one we sketched above.

For graphs of treewidth bounded by some arbitrary but fixed integer $k$, one can solve EQUITABLE COLORING in polynomial time, even when the number $r$ of colors is not constant and given as part of the input (albeit the order of the polynomial depends on $k$). This was recently shown by Bodlaender and Fomin [6] using a combinatorial result of Kostochka, Nakprasit, and Pemmaraju [17].

## 3.2 General Factors

Lovász [19, 20] introduced the following problem.

> GENERAL FACTOR
>
> *Instance:* A graph $G$ and for each vertex $v$ of $G$ a set $K(v) \subseteq \{0, \ldots, d(v)\}$; we call $K(v)$ the *cardinality set* of $v$.
>
> *Question:* Is there a subset $F \subseteq E(G)$ such that for each vertex $v \in V(G)$ the number of edges in $F$ incident with $v$ is an element of $K(v)$?

This problem clearly generalizes the polynomial-time solvable $r$-FACTOR problem where all cardinality sets are equal to $\{r\}$. However, GENERAL FACTOR is easily seen to be NP-hard, already if cardinality sets are restricted to $\{0, 3\}$ and $\{1\}$ (say, by reduction from 3-DIMENSIONAL MATCHING). Cornuéjols [7] gives a full classification of the complexity of GENERAL FACTOR when cardinality sets are restricted to some fixed class of sets (a dichotomy of NP-hard and polynomial-time solvable cases).

**Theorem 3** ([26]). GENERAL FACTOR *is W[1]-hard when parameterized by the treewidth of the instance graph. The problem remains W[1]-hard when the given graph is bipartite and all cardinality sets for vertices of one side of the bipartition are equal to $\{1\}$.*

The proof of this result is, once again, obtained by an fpt-reduction from PARTITIONED CLIQUE.

GENERAL FACTOR can be solved in polynomial time for graphs of bounded treewidth where the order of the polynomial depends on the treewidth bound [29]. In fact, the main result of [29] is a meta-theorem that provides polynomial-time algorithms for a wide range of problems on graphs of bounded treewidth. Each of the covered problems asks for a given graph $G$ with cardinality sets $K(v) \subseteq \{0, \ldots, |V(G)| + |E(G)| - 1\}$ whether there exists a set $X \subseteq V(G) \cup E(G)$ such that

1. for each vertex $v \in V(G)$, the number of vertices in $X$ adjacent to $v$ plus the number of edges in $X$ incident with $v$ belongs to $K(v)$,

2. $X$ satisfies a fixed property $P(X)$ expressible in a certain formalism called "Monadic Second Order Logic."

For example $P(X)$ could state that $X$ is a set of vertices that forms a color class for a proper 3-coloring of $G$. For GENERAL FACTOR the property $P(X)$ just states that $X$ is a set of edges.

## 3.3 Generalized Satisfiability

A *Boolean constraint* is a pair $C = ((x_1, \ldots, x_r), R)$ where $x_1, \ldots, x_r$ are distinct variables and $R \subseteq \{0, 1\}^r$ is a Boolean relation of arity $r > 0$. We write $\text{var}(C) = \{x_1, \ldots, x_r\}$ and say that $C$ *is over* a set $X$ of variables if $\text{var}(C) \subseteq X$. A mapping $\tau : X \to \{0, 1\}$ *satisfies* a Boolean constraint $C = ((x_1, \ldots, x_r), R)$ if $C$ is over $X$ and $(\tau(x_1), \ldots, \tau(x_r)) \in R$.

GENERALIZED SATISFIABILITY

*Instance:* A finite set $X$ of variables and finite set $S$ of Boolean constraints over $X$.

*Question:* Is there a mapping $\tau : X \to \{0, 1\}$ that satisfies all constraints in $S$?

Clearly GENERALIZED SATISFIABILITY is NP-complete, as, for example, it contains 3-SAT as the special case where all constraints use the same relation $R = \{0, 1\}^3 \setminus \{(0, 0, 0)\}$. Schaefer [27] classifies the complexity of GENERALIZED SATISFIABILITY problems for instances that use relations from a fixed class (a dichotomy of NP-hard and polynomial-time solvable cases).

By associating certain graphs to sets of Boolean constraints one can apply the treewidth parameter to the GENERALIZED SATISFIABILITY problem.

Consider an instance $(X, S)$ of GENERALIZED SATISFIABILITY. The *primal graph* has vertex set $X$, two variables are adjacent if they occur together in a constraint. Symmetrically, the *dual graph* has as vertex set $S$, two constraints are adjacent if they share a variable. Finally, the *incidence graph* is the bipartite graph with vertex set $X \cup S$; a constraint and a variable are adjacent if the variable occurs in the constraint.

It is easy to see that GENERALIZED SATISFIABILITY is fixed-parameter tractable if parameterized by the treewidth of primal graphs [14]. However, regarding the treewidth of dual and incidence graphs we have the following negative results.

**Theorem 4** ([24]). GENERALIZED SATISFIABILITY *is W[1]-hard when parameterized by the treewidth of the dual graph or by the treewidth of the incidence graph of the instance.*

We sketch the proof which uses an fpt-reduction from CLIQUE. Let $G$ be a graph with $V(G) = \{v_1, \ldots, v_n\}$. We construct an instance $(X, S)$ of GENERALIZED SATISFIABILITY as follows. First we construct a relation $R \subseteq \{0, 1\}^{2n}$ that encodes the edges of $G$ using Boolean values 0 and 1. For each edge $v_p v_q$ of $G$, $1 \leq p < q \leq n$, we add to $R$ the $2n$-tuple

$$(t_{p,1}, \ldots, t_{p,n}, t_{q,1}, \ldots, t_{q,n})$$

where $t_{p,i} = 1$ if and only if $p = i$, and $t_{q,i} = 1$ if and only if $q = i$, $1 \leq i \leq n$. We let $S$ be the set of Boolean constraints

$$C_{i,j} = ((x_{i,1}, \ldots, x_{i,n}, x_{j,1}, \ldots, x_{j,n}), R)$$

and $X$ the set of variables $x_{i,j}$, for $1 \leq i < j \leq k$. It is easy to verify that $G$ contains a clique on $k$ vertices if and only if $S$ is satisfiable. Since there are exactly $\binom{k}{2}$ constraints in $S$, the treewidth of the dual graph is at most $\binom{n}{k} - 1$, thus bounded in terms of $k$. Using Observation 1 it is easy to see that the treewidth of the incidence graph is at most $\binom{k}{2}$. Hence Theorem 4 follows.

BOOLEAN SATISFIABILITY is defined similarly except that instead of Boolean constraints one considers *clauses* (disjunctions of variables or negated variables). Primal, dual, and incidence graphs are defined for sets of clauses in the obvious way. Interestingly, BOOLEAN SATISFIABILITY and GENERALIZED SATISFIABILITY are of different parameterized complexity: BOOLEAN SATISFIABILITY is fixed-parameter tractable when parameterized by the treewidth of any of the three associated graphs [25, 28].

## 4 A New Hardness Result for the Minimum Maximum Outdegree Problem

A (positive integral) *edge weighting* of a graph $G$ is a mapping $w$ that assigns to each edge of $G$ a positive integer. An *orientation* of $G$ is a mapping $\Lambda : E(G) \to V(G) \times V(G)$ with $\Lambda(uv) \in \{(u,v), (v,u)\}$. The *weighted outdegree* of a vertex $v \in V(G)$ with respect to an edge weighting $w$ and an orientation $\Lambda$ is defined as

$$d^+_{G,w,\Lambda}(v) = \sum_{vu \in E(G) \text{ with } \Lambda(vu)=(v,u)} w(vu).$$

Asahiro, Miyano, and Ono [3] consider the following problem and discuss applications and related problems.

MINIMUM MAXIMUM OUTDEGREE

*Instance:* A graph $G$, an edge weighting $w$ of $G$ given in unary, and a positive integer $r$.

*Question:* Is there an orientation $\Lambda$ of $G$ such that $d^+_{G,w,\Lambda}(v) \leq r$ for each $v \in V(G)$?

We assume that the edge weighting $w$ is given in unary since otherwise the problem is already NP-complete for graphs of treewidth 2, as a simple reduction from PARTITION shows [3]. If all edge weights are identical, then MINIMUM MAXIMUM OUTDEGREE can be solved in polynomial time using network flows [3]. Furthermore, the problem can be solved for graphs of treewidth $k$ in time bounded by a polynomial whose order depends on $k$ [29]. The next theorem shows that this dependence is necessary, unless FPT = W[1].

**Theorem 5.** MINIMUM MAXIMUM OUTDEGREE *is* W[1]-*hard when parameterized by the treewidth of the instance graph.*

*Proof.* We use the following intermediate problem:

CHOSEN MAXIMUM OUTDEGREE

*Instance:* A graph $G$, an edge weighting $w$ of $G$ given in unary, and for each vertex $v \in V(G)$ a non-negative integer $\rho(v)$.

*Question:* Is there an orientation $\Lambda$ of $G$ such that $d^+_{G,w,\Lambda}(v) \leq \rho(v)$ for each $v \in V(G)$? We call such an orientation $\rho$-*admissible*.

**Claim 1.** CHOSEN MAXIMUM OUTDEGREE *fpt-reduces to* MINIMUM MAXIMUM OUTDEGREE *(both problems are parameterized by the treewidth of the instance graph).*

To prove this claim, let $G, w, \rho$ be an instance of CHOSEN MAXIMUM OUTDEGREE. We construct an edge weighted graph $H$ from $G$ as follows. Let $r = \max_{v \in V(G)} \rho(v)$. For each vertex $v \in V(G)$ with $\rho(v) < r$ we add to $G$ two new vertices $x_v, y_v$ and the edges $vx_v$, $vy_v$, and $x_v y_v$ with edge weights $r - \rho(v)$, $r - \rho(v)$, and $r$, respectively. It is easy to verify that $H$ has an orientation with maximum weighted outdegree at most $r$ if and only if $G$ has a $\rho$-admissible orientation. Thus Claim 1 follows.

Next we give an fpt-reduction from PARTITIONED CLIQUE to CHOSEN MAXIMUM OUTDEGREE; the theorem will then follow by Claim 1.

Consider a $k$-partite graph $G$ with partition $V_1, \ldots, V_k$ with $|V_1| = \cdots = |V_k| = n$. We write $V_i = \{v_i^1, \ldots, v_i^n\}$ for $1 \leq i \leq k$. For $1 \leq i < i' \leq k$ let $E_{i,i'} = \{(q, q') : 1 \leq q \leq n, 1 \leq q' \leq n, v_i^q v_{i'}^{q'} \in E(G)\}$. We are going to construct a graph $H$ with edge weighting $w$ and vertex weighting $\rho$.

The vertex set of $H$ is obtained as follows:

1. For $1 \leq i \leq k$ and $1 \leq j \leq n$, we add to $V(H)$ three vertices $u_i^j$, $x_i^j$, and $y_i^j$.

2. For $1 \leq i \leq k$ we add to $V(H)$ a vertex $a_i$.

3. For $1 \leq i < i' \leq k$ we add to $V(H)$ vertices $b_{i,i'}$, $c_{i,i'}$, and $d_{i,i'}$.

4. For $1 \leq i < i' \leq k$ and each $(q, q') \in E_{i,i'}$ we add to $V(H)$ a vertex $e_{i,i'}^{q,q'}$.

The edge set of $H$ is obtained as follows.

1. For $1 \leq i \leq k$ and $1 \leq j \leq n$ we add the edges $a_i u_i^j$, $u_i^j x_i^j$, and $u_i^j y_i^j$.

2. For $1 \leq i < i' \leq k$ and $(q, q') \in E_{i,i'}$ we add the edges $e_{i,i'}^{q,q'} d_{i,i'}$, $e_{i,i'}^{q,q'} b_{i,i'}$ and $e_{i,i'}^{q,q'} c_{i,i'}$.

3. For $1 \leq i < i' \leq k$, $1 \leq j \leq n$, and $1 \leq j' \leq n$ we add the edges $x_i^j b_{i,i'}$, $y_i^j c_{i,i'}$, and $x_{i'}^{j'} b_{i,i'}$, $y_{i'}^{j'} c_{i,i'}$.

We shall refer to the edges added in the last step as *special edges*.

**Claim 2.** *The treewidth of $H$ is at most $2\binom{k}{2} + 1$.*

Indeed, the set $BC = \{b_{i,i'}, c_{i,i'} : 1 \leq i < i' \leq k\}$ is of cardinality $2\binom{k}{2}$ and $H - BC$ is a disjoint union of trees. Hence Claim 2 follows from Observation 1.

Let $N = n + 1$. We define the weights of special edges:

$$w(x_i^j b_{i,i'}) = w(y_i^j c_{i,i'}) = N^3 + j \quad (i < i',\ 1 \leq j \leq n)$$
$$w(x_{i'}^j b_{i,i'}) = w(y_{i'}^j c_{i,i'}) = N^3 + jN \quad (i < i',\ 1 \leq j \leq n)$$

Let $M(v)$ denote the sum of the weights of all special edges incident with vertex $v$. We set $M = k(N^3 + N^2)$ to ensure that we have $M(x_i^j) < M(y_i^j) < M$ for all $1 \leq i \leq k$ and $1 \leq j \leq n$.

We define further edge and vertex weights as follows.

For $1 \leq i \leq k$ we set $\rho(a_i) = 1$ and $w(a_i u_i^j) = 1$, for $1 \leq j \leq n$.

For $1 \leq i \leq k$ and $1 \leq i \leq n$ we set $w(u_i^j x_i^j) = \rho(x_i^j) = M$ and $w(u_i^j y_i^j) = \rho(u_i^j) = \rho(y_i^j) = M + 1$.

For $1 \leq i < i \leq k$ we set $\rho(d_{i,i'}) = |E_{i,i'}| - 1$, and for $(q, q') \in E_{i,i'}$ we define:

$$w(d_{i,i'} e_{i,i'}^{q,q'}) = 1$$
$$w(e_{i,i'}^{q,q'} b_{i,i'}) = w(x_i^q b_{i,i'}) + w(x_{i'}^{q'} b_{i,i'})$$
$$w(e_{i,i'}^{q,q'} c_{i,i'}) = w(y_i^q c_{i,i'}) + w(y_{i'}^{q'} c_{i,i'})$$
$$\rho(e_{i,i'}^{q,q'}) = w(e_{i,i'}^{q,q'} c_{i,i'}) \quad (> w(e_{i,i'}^{q,q'} b_{i,i'}))$$

For $1 \leq i < i \leq k$ we define:

$$\rho(b_{i,i'}) = \sum_{j=1}^n w(x_i^j b_{i,i'}) + \sum_{j=1}^n w(x_{i'}^j b_{i,i'})$$

$$\rho(c_{i,i'}) = \sum_{j=1}^n w(y_i^j c_{i,i'}) + \sum_{j=1}^n w(y_{i'}^j c_{i,i'})$$

**Claim 3.** *If $H$ has a $\rho$-admissible orientation then $G$ has a clique on $k$ vertices.*

To prove this claim, let $\Gamma$ be an admissible orientation. Let $A = \{\Lambda(e) : e \in E(G)\}$. We shall use terminology for directed graphs. For example, if $(x, y) \in A$ then we say that $xy$ is an "outgoing edge" of $x$ and an "incoming edge" of $y$.

Let $1 \leq i \leq k$. Observe that $a_i$ has at most one outgoing edge. If it has no outgoing edges, then we can revert an arbitrarily chosen one maintaining a $\rho$-admissible orientation. Hence, without loss of generality, we may assume that $a_i$ has exactly one outgoing edge, say $(a_i, u_i^{p(i)}) \in A$ for some $p(i) \in \{1, \ldots, n\}$. Consequently, for all $j \in \{1, \ldots, n\} \setminus \{p(i)\}$ we have $(y_i^j, u_i^j) \in A$, and in turn $(c_{i',i}, y_i^j) \in A$ for all $1 \leq i' \leq k$.

Let $1 \leq i < i' \leq k$. For similar reasons as in the previous paragraph we may assume, without loss of generality, that $d_{i,i'}$ has exactly one incoming edge, say $(e_{i,i'}^{q(i,i'),q'(i,i')} d_{i,i'}) \in A$ for $(q(i, i'), q'(i, i')) \in E_{i,i'}$. It follows that $(c_{i,i'}, e_{i,i'}^{q(i,i'),q'(i,i')}) \in A$. We have already concluded that $(c_{i,i'}, y_i^j) \in A$ for all $j \in \{1, \ldots, n\} \setminus \{p(i)\}$ and $(c_{i,i'}, y_{i'}^j) \in A$ for all $j \in \{1, \ldots, n\} \setminus \{p(i')\}$. Thus the number of outgoing edges from $c_{i,i'}$ is at least $2(n-1) + 1 = 2n - 1$. Observe that each edge incident with $c_{i,i'}$ has weight greater than $N^3$, the weight of $c_{i,i'} e_{i,i'}^{q(i,i'),q'(i,i')}$ is even greater than $2N^3$. Since $\lfloor \rho(c_{i,i'})/N^3 \rfloor = 2n$, we conclude that $c_{i,i'}$ has no further outgoing edges than the $2n - 1$ edges identified so far. In particular $(e_{i,i'}^{q,q'}, c_{i,i'}) \in A$ for all $(q, q') \in E_{i,i'} \setminus \{(q(i, i'), q'(i, i'))\}$ and $(y_i^{p(i)} c_{i,i'}), (y_{i'}^{p(i')} c_{i,i'}) \in A$. The latter implies $(u_i^{p(i)}, y_i^{p(i)}) \in A$ and consequently $(x_i^{p(i)}, u_i^{p(i)}) \in A$ and $(b_{i,i'}, x_i^{p(i)}) \in A$; similarly $(b_{i,i'}, x_{i'}^{p'(i)}) \in A$. We concluded above that for $(q, q') \in E_{i,i'} \setminus \{(q(i, i'), q'(i, i'))\}$ we have $(e_{i,i'}^{q,q'}, c_{i,i'}) \in A$, thus $(b_{i,i'}, e_{i,i'}^{q,q'}) \in A$. Hence the weighted outdegree of $b_{i,i'}$ is high enough to conclude, similarly as above for $c_{i,i'}$, that all edges incident with $b_{i,i'}$ that we have not yet identified as outgoing are incoming edges.

In view of $\rho(b_{i,i'})$ and $\rho(c_{i,i'})$ and the weights of the respective outgoing edges we conclude

$$w(b_{i,i'} x_i^{q(i,i')}) + w(b_{i,i'} x_{i'}^{q'(i,i')}) = w(b_{i,i'} e_{i,i'}^{q(i,i'),q'(i,i')})$$
$$\geq w(b_{i,i'} x_i^{p(i)}) + w(b_{i,i'} x_{i'}^{p(i')})$$

and

$$w(c_{i,i'} y_i^{q(i,i')}) + w(c_{i,i'} y_{i'}^{q'(i,i')}) = w(c_{i,i'} e_{i,i'}^{q(i,i'),q'(i,i')})$$
$$\leq w(c_{i,i'} y_i^{p(i)}) + w(c_{i,i'} y_{i'}^{p(i')}).$$

The first inequality gives $q(i, i') + q'(i, i')N \geq p(i) + Np(i')$ and so $q'(i, i') \geq p(i')$; the second inequality gives $q'(i, i') \leq p(i')$; their combination gives $q'(i, i') = p(i')$. Using this identity to simplify the two inequalities we can finally obtain $q(i, i') = p(i)$. We conclude that $v_i^{p(i)}$ and $v_{i'}^{p(i')}$ are adjacent in $G$ for all $1 \leq i < i' \leq k$. Consequently the vertices $v_1^{p(1)}, \ldots, v_k^{p(k)}$ induce a clique in $G$, and Claim 3 follows.

**Claim 4.** *If $G$ has a clique on $k$ vertices then $H$ has a $\rho$-admissible orientation.*

This is the easy direction. Assume there exists a clique on $k$ vertices in $G$. We can write the vertices of the clique as $v_1^{p(1)}, \ldots, v_k^{p(k)}$ where $p(i) \in \{1, \ldots, n\}$. Clearly $(p(i), p(i')) \in E_{i,i'}$ holds for $1 \leq i < i' \leq k$. We define a $\rho$-admissible orientation $\Lambda$ es follows. Again we write $A = \{\Lambda(e) : e \in E(G)\}$.

For $1 \leq i \leq k$ we make $a_i u_i^{p(i)}$ the only outgoing edge of $a_i$; accordingly for $j = p(i)$ we set $(u_i^j, y_i^j), (x_i^j, u_i^j) \in A$, and $(y_i^j, c), (c, x_i^j) \in A$ for all $c = c_{i',i}$ ($1 \leq i' < i$) and $c = c_{i,i'}$ ($i < i' \leq k$); for $j \neq p(i)$ we take the inverse orientation of the mentioned edges.

For $1 \leq i < i' \leq k$ we we make $d_{i,i'} e_{i,i'}^{p(i),p(i')}$ the only incoming edge of $d_{i,i'}$; for $(p, p') \in E_{i,i'}$ we set $(e_{i,i'}^{p,p'}, b_{i,i'}) \in A$ exactly when $j = p(i)$, and we set $(e_{i,i'}^{p,p'}, c_{i,i'}) \in A$ exactly when $j \neq p(i)$.

This completes the definition of $\Lambda$. It is easy to verify that $\Gamma$ is indeed $\rho$-admissible, hence Claim 4 follows. □

It is evident that $H$ can be computed in polynomial time from $G$. By Claim 2, the treewidth of $H$ is a function of $k$, thus with Claims 3 and 4 we have established an fpt-reduction from PARTITIONED CLIQUE to CHOSEN MAXIMUM OUTDEGREE. In view of Claim 1 and the W[1]-hardness of PARTITIONED CLIQUE, Theorem 5 follows.

## Acknowledgement

Research partially supported by the European Research Council, grant reference 239962.